\documentclass[lettersize,journal]{IEEEtran}
\usepackage{amsmath,amsfonts}
\usepackage{array}
\usepackage[caption=false,font=normalsize,labelfont=sf,textfont=sf]{subfig}
\usepackage{textcomp}
\usepackage{stfloats}
\usepackage{url}
\usepackage{verbatim}
\usepackage{graphicx}
\usepackage{cite}
\usepackage{balance}

\usepackage{graphicx}

\newcommand{\para}[1]{\vspace{2pt}\noindent\textbf{#1.~}}
\newcommand{\ignore}[1]{}
\newcommand{\system}{\sloppy{Satellite\@}}

\usepackage{algorithm}

\newcommand{\vuln}{\sloppy{SMV\@}}
\newcommand{\bugname}{{\vuln}}

\usepackage{ulem}
\usepackage[switch]{lineno}
\newcommand{\publicUrl}{\url{https://github.com/Janelinux/satellite.git}}
\usepackage{listings, xcolor}
\definecolor{verylightgray}{rgb}{.97,.97,.97}
\lstdefinelanguage{Solidity}{
  keywords=[1]{anonymous, assembly, assert, balance, break, call, callcode, case, catch, class, constant, continue, constructor, contract, debugger, default, delegatecall, delete, do, else, emit, event, experimental, export, external, false, finally, for, function, gas, if, implements, import, in, indexed, instanceof, interface, internal, is, length, library, log0, log1, log2, log3, log4, memory, modifier, new, payable, pragma, private, protected, public, pure, push, require, return, returns, revert, selfdestruct, send, solidity, storage, struct, suicide, super, switch, then, this, throw, transfer, true, try, typeof, using, value, view, while, with, addmod, ecrecover, keccak256, mulmod, ripemd160, sha256, sha3}, 
  keywordstyle=[1]\color{blue}\bfseries,
  keywords=[2]{address, bool, byte, bytes, bytes1, bytes2, bytes3, bytes4, bytes5, bytes6, bytes7, bytes8, bytes9, bytes10, bytes11, bytes12, bytes13, bytes14, bytes15, bytes16, bytes17, bytes18, bytes19, bytes20, bytes21, bytes22, bytes23, bytes24, bytes25, bytes26, bytes27, bytes28, bytes29, bytes30, bytes31, bytes32, enum, int, int8, int16, int24, int32, int40, int48, int56, int64, int72, int80, int88, int96, int104, int112, int120, int128, int136, int144, int152, int160, int168, int176, int184, int192, int200, int208, int216, int224, int232, int240, int248, int256, mapping, string, uint, uint8, uint16, uint24, uint32, uint40, uint48, uint56, uint64, uint72, uint80, uint88, uint96, uint104, uint112, uint120, uint128, uint136, uint144, uint152, uint160, uint168, uint176, uint184, uint192, uint200, uint208, uint216, uint224, uint232, uint240, uint248, uint256, var, void, ether, finney, szabo, wei, days, hours, minutes, seconds, weeks, years},  
  keywordstyle=[2]\color{teal}\bfseries,
  keywords=[3]{block, blockhash, coinbase, difficulty, gaslimit, number, timestamp, msg, data, gas, sender, sig, value, now, tx, gasprice, origin},  
  keywordstyle=[3]\color{violet}\bfseries,
  identifierstyle=\color{black},
  sensitive=false,
  comment=[l]{//},
  morecomment=[s]{/*}{*/},
  commentstyle=\color{gray}\ttfamily,
  stringstyle=\color{red}\ttfamily,
  morestring=[b]',
  morestring=[b]"
}
\usepackage{booktabs}
\usepackage{oplotsymbl}
\usepackage{subfig}
\usepackage{siunitx}
\usepackage{array,framed}
 \usepackage{
   color,
   float,
   epsfig,
   wrapfig,
   graphics,
   graphicx,
   subcaption
 }
\usepackage{setspace}
\usepackage{amsfonts}
\usepackage{latexsym,fancyhdr,url}
\usepackage{enumerate}
\usepackage{algpseudocode}
\usepackage{graphics}
\usepackage{xparse} 
\usepackage{xspace}
\usepackage{multirow}
\usepackage{microtype}
\usepackage{graphicx}
\usepackage{pifont}
\usepackage{wasysym}   
\usepackage{xcolor}    

\usepackage{fancyhdr}

\begin{document}

\title{\system{}: Detecting and Analyzing Smart Contract Vulnerabilities caused by Subcontract Misuse}

\author{
Zeqin~Liao, 
Yuhong~Nan,~\IEEEmembership{Member,~IEEE}, 
Zixu~Gao, 
Henglong~Liang, 
Sicheng~Hao, 
Jiajing~Wu, 
and~Zibin~Zheng,~\IEEEmembership{Fellow,~IEEE}         

\thanks{ 
\setlength{\parindent}{0pt}
Zeqin~Liao, Yuhong~Nan, Zixu~Gao, Henglong~Liang, Sicheng~Hao, Jiajing~Wu and~Zibin~Zheng are with School of Software Engineering, Sun Yat-sen University, China. 

E-mail: \{liaozq8, gaozx9, lianghlong, haosch\} @mail2.sysu.edu.cn

E-mail: \{nanyh, wujiajing, zhzibin\}@mail.sysu.edu.cn } 

\thanks{
\setlength{\parindent}{0pt}
Yuhong Nan is the corresponding author}


}


\markboth{Journal of \LaTeX\ Class Files,~Vol.~14, No.~8, August~2021}%
{Shell \MakeLowercase{\textit{et al.}}: A Sample Article Using IEEEtran.cls for IEEE Journals}


\maketitle

\begin{abstract}
Code reuse is a common practice in software engineering. Developers of smart contracts pervasively reuse subcontracts to improve development efficiency. Like any program language, such subcontract reuse may unexpectedly include, or introduce vulnerabilities to the end-point smart contract. Indeed, prior empirical studies have identified a number of issues caused by code reuse in smart contracts. Unfortunately, automatically detecting such issues poses several unique challenges. Particularly, in most cases, smart contracts are compiled as bytecode, whose class-level information (e.g., inheritance, virtual function table), and even semantics (e.g., control flow and data flow) are fully obscured as a single smart contract after compilation. Therefore, it is rather difficult to identify the reused parts of subcontract from a given smart contract, not to mention finding potential vulnerabilities caused by subcontract misuse.


In this paper, we propose \system{}, a new bytecode-level static analysis framework for subcontract misuse vulnerability (SMV) detection in smart contracts. \system{} incorporates a series of novel designs to enhance its overall effectiveness..  Particularly, \system{} utilizes a transfer learning method to recover the inherited methods, which are critical for identifying subcontract reuse in smart contracts. Further, \system{} extracts a set of fine-grained method-level features and performs a method-level comparison, for identifying the reuse part of subcontract in smart contracts. Finally, \system{} summarizes a set of \bugname{} indicators according to their types, and hence effectively identifies \bugname{}s.  To evaluate \system{}, we construct a dataset consisting of 58 SMVs derived from real-world attacks and collect additional 56 SMV patterns from SOTA studies. 
Experiment results indicate that \system{} exhibits good performance in identifying SMV, with a precision rate of 84.68\% and a recall rate of 92.11\%.
In addition, \system{} successfully identifies 14 new/unknown \bugname{} over 10,011 real-world smart contracts, affecting a total amount of digital assets worth 201,358 USD.
\end{abstract}

\begin{IEEEkeywords}
Smart Contract, Static Analysis, Subcontract Misuse, Vulnerability Detecion.
\end{IEEEkeywords}

\section{Introduction}
\label{sec:intro}

The smart contract is a specific type of program that executes on the blockchain. 
Solidity is one of the mainstream languages employed in the smart contract programming, which is extensively applied to major blockchain platforms, including Ethereum~\cite{Ethereum}, TRON~\cite{TRON}, and BNB Chain~\cite{BNB}.
Smart contracts 
are foundational to the functioning of diverse decentralized applications (DApps) across blockchain ecosystems, including decentralized finance (DeFi) and decentralized governance.


A key factor contributing to the success of smart contracts is the presence of various well-audited subcontracts, such as ERC20 standard subcontract. These subcontracts offer reusable functionality, enabling developers to efficiently and securely create smart contracts without the need to "reinvent the wheel". However,
the inappropriate use of these subcontracts by developers(e.g., reusing subcontracts that conflict with one another) can render the reuse of these subcontracts unsafe. Given that smart contracts often manage substantial digital assets, adversaries can exploit vulnerabilities arising from the misuse of subcontracts, leading to significant economic losses.
In our research, we refer to such vulnerabilities in smart contracts as the Subcontract Misuse Vulnerability (\bugname{}). Our investigation shows that,
the causes of SMV lie in two aspects: (1) Variable conflict between main contracts and subcontracts
or between subcontracts; and (2) lack of security check due to misunderstanding their
capacity.
As of July 2024, smart contracts have suffered more than 16 \bugname{} security incidents, with an accumulated loss of 566 million USD\footnote{
The sources of incident
statistics are listed in 
\url{https://figshare.com/s/3f307813081b7888edba?file=45237472}}.



Given the severe loss caused by \bugname{}, there is still limited research that aims at analyzing \bugname{}. 
Specifically, previous studies~\cite{Huang2024Reveal,sun2023Demystifying} conduct empirical analysis to understand various patterns of subcontract reuse.
Based on their observations, these studies offer a series of recommendations on how to reuse subcontracts effectively. However, such recommendations do not adequately address the security issues in smart contracts, particularly with respect to SMV detection. 
Additionally, a certain number of studies have been devoted to code clone identification,
which involves comparing the similarity between a given smart contract and known vulnerable
contracts to detect code clones. However, since \bugname{} detection requires the fine-grained analysis
on data flow (e.g., variable conflict) and control flow (e.g., access control), methodologies developed
for code clone detection are insufficiently capable for SMV detection.
Moreover, another notable effort related to our domain is ZepScope [27]. ZepScope primarily focuses on identifying inconsistencies in the implementation of security checks between official libraries and their copies within smart contracts, which are actually a subset of \bugname. As a result, ZepScope lacks the generalization and capability to support SMV detection.
Furthermore, most of these works are insufficiently generic, because they require analyzing the source code of smart contracts, which is impracticable for most smart contracts  (i.e., only bytecode is accessible).


\para{Our work} 
In this paper, we propose a novel bytecode-level static analysis framework, \system{}, to identify \bugname{} for smart contracts. To the best of our knowledge, \system{} is the first program analysis approach that identifies \bugname{} based on the contract bytecode. Using this approach, \system{} can automatically perform security audits on various smart contract applications before deployment, thereby mitigating potential risks that could lead to severe damage, such as economic losses.


Unlike traditional languages such as Java and C++, subcontract information (i.e., class structure of smart contract) are entirely absent in Solidity contract bytecode~\cite{Solidity}.
The key challenge in this research is to 
establish a method-level similarity comparison for identifying subcontract usage within the smart contract at the bytecode level, 
which is crucial for determining which cause of SMV (e.g., variable conflict or access control) could be exploited by an adversary.
However, the state-of-the-art (SOTA) decompilers (e.g., Gigahorse~\cite{grech2019Gigahorse}) fail to recover  inherited methods. This limitation poses a significant obstacle to SMV detection since inheritance is a primary mechanism through which smart contracts reuse subcontracts. 
Besides,  Previous studies~\cite{zhang2019libid, wu2023libscan}  extract the limited features (e.g., method call, field, control flow) for identifying code reuse. These features are insufficient for identifying SMVs, as SMV detection requires a more comprehensive analysis that encompasses the unique characteristics of smart contracts such as Gas~\cite{Solidity}.


To identify \bugname{} effectively, \system{} features the following novel designs. 
Firstly, Satellite leverages transfer learning to accurately recover inherited methods from the bytecode of smart contracts by learning patterns that define the boundaries (i.e., start and end points) of these methods from a training corpus (see Section~\ref{sec: RecoveryforInheritance}). 
While inherited methods are critical for \bugname{} detection, they are totally lost during the compilation process from source code to bytecode. 
Secondly, \system{} complements and extends previous research by considering four new types of contract-specific features, including new mechanisms, message calls, events, and pre-compiled methods(see Section~\ref{sec: SignatureGeneration}).
And Satellite captures these features across methods and conducts precise method-level similarity comparisons to identify the usage of subcontracts within a given contract (see Section~\ref{sec: ReusageIdentification}).
Further, we analyze the official subcontract functions and attack reports to derive
apriori knowledge, including access control specifications and conflicts of subcontracts. 
 Subsequently, \system{} highlights the apriori knowledge as the \bugname{} indicators to support vulnerability detection. Finally, \system{} employs taint analysis to determine whether an external contract can be invoked and reach state variables associated with the identified \bugname{} indicators. 
Therefore, \system{} effectively identifies \bugname{}
based on the identified subcontract usages (see Section~\ref{sec: VulnerabilityIdentification}).





To evaluate the effectiveness of \system{}, we exhaustively search for SMV attack reports from the public, construct a manually-labeled attack dataset with a groundtruth of 58 \bugname{} instances. 
Further, we extract additional 56 contracts containing SMVs from SOTA studies and open-source platforms (e.g., GitHub).
Our evaluation results demonstrate that \system{} achieves the SMV detection with a precision rate of 84.68\% and a recall rate of 92.11\%. These findings indicate that \system{} effectively detects the majority of \bugname{}s that cause real-world damage.
To further evaluate the impact of \bugname{} in the wild, we apply Satellite over 10,011 real-world smart contracts randomly selected from the largest open-source smart contract dataset ~\cite{xblockcontract}. The large-scale analysis results reveal that \system{} identifies 14 new SMVs that have never been detected by previous research, collectively affecting digital assets valued at 201,358 USD. 


In this paper, we make the following contributions.

\begin{itemize}
    \item We highlight the cause of SMV, including variable conflict and lack of security check.

    \item We propose \system{}, the first bytecode-level static analysis framework for detecting subcontract-misuse vulnerability in smart contract. 
    \item  We conduct an in-depth evaluation to show the effectiveness of \system{}. Particularly, \system{} identifies 14 new \bugname{}s over 10,011 real-world smart contracts. 
    \item  
    We release the artifact of \system{}, as well as the datasets in \publicUrl.
\end{itemize}


\section{Background and Motivation}
\label{sec:background}

\subsection{Smart Contract and Code Reuse}
\label{sec: ThirdPartyLibrary}

Smart contract is a specific type of program that executes on the blockchain. Solidity is one of the most popular languages for smart contracts. Written in Solidity source code, smart contracts are compiled into bytecode. 
Structurally, a smart contract is composed of a group of methods and variables. Contract state refers to the persistent data that is read and written through global variables (i.e., state variables). Owing to the expensive storage space of blockchain, state variables are utilized to store critical content of smart contracts such as the users’ token balance. Therefore, if an adversary can exploit the vulnerability to manipulate critical state variables, smart contracts may suffer severe and undesirable effects (e.g., economic loss).



Code reuse is a common practice in software engineering that involves developing new programs by utilizing pre-existing code. Unlike traditional software such as Android applications, most smart contract developers avoid reusing code from untrusted sources (e.g., web pages) because of the high-value assets typically managed within smart contracts. Consequently, the predominant method of code reuse in the context of smart contracts involves inheriting from trusted subcontracts. These subcontracts can be categorized into four types: Interface, Abstract, Library, and Executable subcontract. Each type serves a distinct role, ensuring that the codebase remains secure while facilitating the modular and efficient development of smart contract. 

According to Solidity documentation~\cite{Solidity}, the inheritance system in Solidity is similar to that of Python. However, unlike Python, due to the demand of economizing Gas (i.e., execution fuel of smart contract) and storage space, contract bytecode does not contain the information of the class (e.g., subcontract). Specifically, when a contract inherits from other base subcontracts, only the single main contract is created on the blockchain, and the code from all base subcontracts is compiled into the created contract, without preserving any class and method information for subcontracts.

\begin{figure*}[t]
\centering
\includegraphics[width=6.5in]{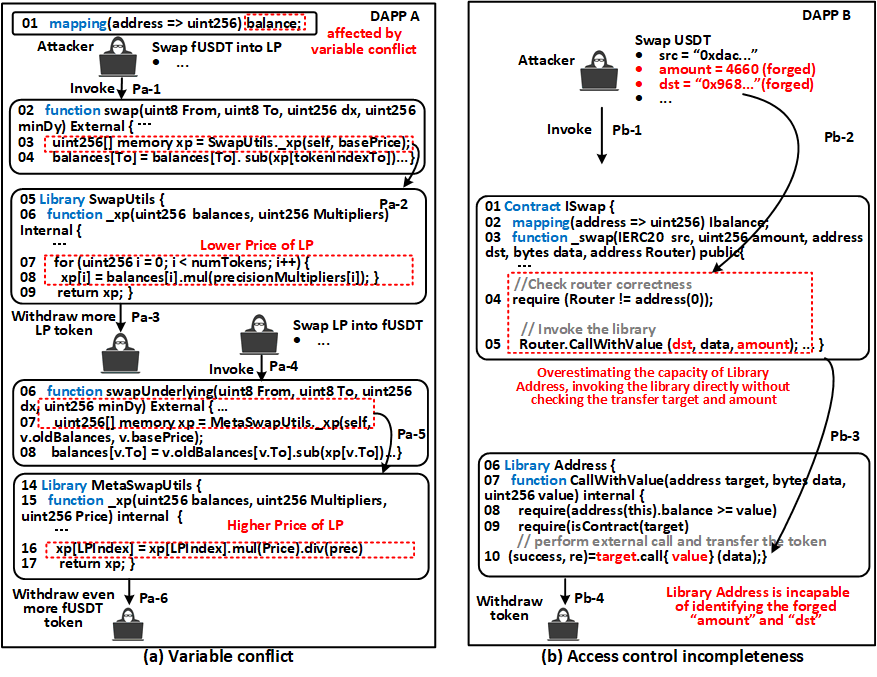}

\caption{Two motivating examples of subcontract misuse vulnerability. }

\label{motivatingexample}
\end{figure*}

\subsection{Definition and Problem Statement}
\label{sec: problemstatement}

\para{Subcontract Misuse Vulnerability} 
According to the collected 16 security incidents, we summarize two types of the \bugname{}. In Figure 1, we utilize two motivating examples to illustrate \bugname{}. The examples are gathered from two real-world smart contracts exploited by adversaries (i.e.,  Nerve~\cite{Nerve}, and Rabby Swap~\cite{RabbySwap}). To better illustrate, we reorganized the original contract code.

\para{Type-1. Variable conflict between smart contract and subcontract} Such type of \bugname{} is due to the issue that, functionally similar methods that modify the critical variables (e.g., token value), are implemented inconsistently between subcontracts or between subcontract and smart contracts. Figure~\ref{motivatingexample} presents an example of \bugname{} caused by the variable conflict between subcontracts. 
Specifically, \textit{xp} is a critical variable for token swap in \textit{DApp A}, which is used for providing the valuation of each token. As can be seen, two libraries reused by \textit{DApp A} (i.e., Library \textit{SwapUtils} and \textit{MetaSwapUtils}) differently implement methods \textit{\_xp}. Further, the calculation of \textit{xp} is implemented differently between the two libraries (line 07-08 and 15), which allows an adversary to leverage the discrepancy to exploit for profit gain. 
As shown in path $P_a$, an adversary first invokes method \textit{swap} to exchange fUSDT token into LP token ($P_{a-1}$). Since method \textit{swap} invokes method \textit{\_xp} that provides a lower price for LP token ($P_{a-2}$), the adversary is capable of purchasing more LP token at a price below the market rate ($P_{a-3}$).
Further, an adversary subsequently invokes method \textit{swapUnderlying} to exchange LP token back into fUSDT token ($P_{a-4}$). Since method \textit{swapUnderlying} invokes method \textit{\_xp} which provides a higher price for LP token ($P_{a-5}$),  allowing the adversary to exchange the LP tokens at a price above the market rate and withdraw more fUSDT tokens ($P_{a-6}$).


\para{Type-2. Lack of security check} 
Smart contracts may omit critical security
checks or implement incorrect access control mechanisms when reusing subcontract, due to contract developer's misunderstanding of the subcontract capacities.
Figure~\ref{motivatingexample}(b) shows an instance of \bugname{} caused by Lack of security check. As can be seen, contract \textit{Swap} utilizes method \textit{\_swap} for the exchange of tokens. Method \textit{CallwithValue} of Library \textit{Address} is an internal method that cannot be accessed by external calls and relies on arguments passed by trusted internal calls for token transfers (line 10).
Unfortunately, as an external method, method \textit{\_swap} overestimates the capacity of Library \textit{Address}, and invokes the library directly without validating the arguments(i.e., \textit{dst} and \textit{amount}) that it passed into the internal method \textit{CallwithValue}
(line 04-05). Therefore, this access control omission introduces an exploitable \bugname{}. 
As presented in path $P_b$, to attack this smart contract, an adversary firstly forge the exchange amount(i.e., \textit{amount}) and destination address (i.e., \textit{dst})  arbitrarily ($P_{b-1}$) and bypass the security checks due to the lack of check on these arguments ($P_{b-2}$), and subsequently the forged exchange amount and destination address are directly used for external transfer ($P_{b-3}$), and finally
withdraw the asset ($P_{b-4}$).
Actually, this vulnerability can be mitigated by adding checks
such as $"require (Ibalance[dst] > amount)"$ between line 4 and 5.





\para{Existing work and limitations}
Despite the widespread exploitation of \bugname{}, there are still limited studies on detecting such types of vulnerability.
To the best of our knowledge, the most pertinent studies for \bugname{} detection comprise two empirical studies. 
Huang et.al~\cite{Huang2024Reveal} analyze library reuse through an empirical analysis of audit reports from various auditing companies. Sun et.al~\cite{sun2023Demystifying} explore subcontract reuse by examining reuse purpose, frequency, functional properties, and development patterns. While these studies provide valuable insights into code reuse patterns, they do not address the detection of SMVs directly. Additionally, a certain number of works~\cite{samreen2022volcano, gao2021Checking, liu2019enabling} have been dedicated to code clone identification, which involves comparing the similarity between a given smart contract and known vulnerable contracts to identify code clones.
Since SMV detection relies heavily on the fine-grained analysis on data flow (e.g., variable conflict) and control flow (e.g., access control),  methodologies developed for code clone detection are insufficiently capable for SMV detection.
Lastly, another notable effort related to \bugname{} detection is ZepScope~\cite{liu2024using}, which investigates security checks inconsistencies between the official OpenZeppelin library and its copies within applications. 
Nevertheless, ZepScope lacks generalizability to support \bugname{} detection due to its focus on library consistency—a relatively small subset of \bugname{} concerns.

Another key limitation of most of the existing studies is their lack of generalizability. Since these studies operate predominantly at the source code level, they only address a limited subset of smart contracts. A recent report~\cite{ContractStatistic} highlights that over 99\% of Ethereum smart contracts do not publicly disclose their source code. 
Given this context, while SMVs are increasingly  becoming a significant threat, there is a lack of an effective framework that can detect SMVs in a generic manner.



\para{Scope of this study} Satellite’s scope covers lack of security check and variable conflict for the following reasons: (1) Severity of the two types of SMVs. As of July 2024, smart contracts have suffered more than 16 security incidents caused by these two types of SMVs, with an accumulated loss of 566 million USD; (2) Representativeness of the SMV types. As clarified by prior research [4] [5], each of these two SMV types is one of the most representative and prevalent SMV types; (3)Under-addressed nature. As discussed above, none of the existing studies can support the detection on both these two types of SMVs.

%



\section{Design of \system{}}
\label{sec:overview}




As mentioned earlier, the root cause of \bugname{} stems from the inadvertent misuse errors within smart contracts, resulting in the manipulation of state variables by an adversary.
To effectively identify SMVs, a straightforward idea for identifying \bugname{} includes the two key steps, i.e., (1) Identify the usages of subcontract via method-level similarity comparison, and (2) Identify SMV through fine-grained control-flow and data-flow analysis based on the impact of such subcontract usages.

\begin{enumerate}[(1).]
\item \textbf{Identify subcontract usage.} Similar to prior research for code clone detection~\cite{gao2021Checking, liu2019enabling}, we can identify which version and methods of subcontract are reused by a given smart contract, based on code similarity analysis. Turning to the example shown in Figure~\ref{motivatingexample} (a), by analyzing the similarity between the official libraries (i.e., \textit{SwapUtils} and \textit{MetaSwapUtils}) and their copies within \textit{DApp A}, we can pinpoint that \textit{DApp A} reuses the method \textit{\_xp} of Library \textit{SwapUtils} as well as the method \textit{\_xp} of Library \textit{MetaSwapUtils} respectively.

\item\textbf{Identify \bugname{}.}
We analyze the official subcontract functionality and attack reports to derive apriori knowledge, such as access control specifications and conflicts of subcontracts.  
For a specific security-sensitive state variable, we can check whether there is a valid path (call chain) along the control flow and data flow, allowing an external call exploit subcontract misuse to arbitrarily manipulate the state variable. Again, take  the example shown in Figure~\ref{motivatingexample} (a), according to the collected apriori knowledge, the method \textit{\_xp} of Library \textit{SwapUtils} exists functionality overlap with method \textit{\_xp} of Library \textit{MetaSwapUtils}. Subsequently,
for the state variable \textit{balance} modified in method \textit{swap} and \textit{swapUnderlying} (line 04 and 12), it can
be actually affected by the adversary due to the variable conflict between two \textit{\_xp} method of  Library \textit{SwapUtils} and \textit{MetaSwapUtils}. To this end, we
identified that the contract comprises a Subcontract-misuse
vulnerability

\end{enumerate}

\subsection{Challenges and Solutions}
\label{sec: challenges}

Due to the inherent and unique characteristics of Solidity, establishing a method-level similarity comparison framework on the bytecode level is by no means trivial for smart contract,  because of the following reasons.


\begin{figure*}[t]
\centering
\includegraphics[width=7in]{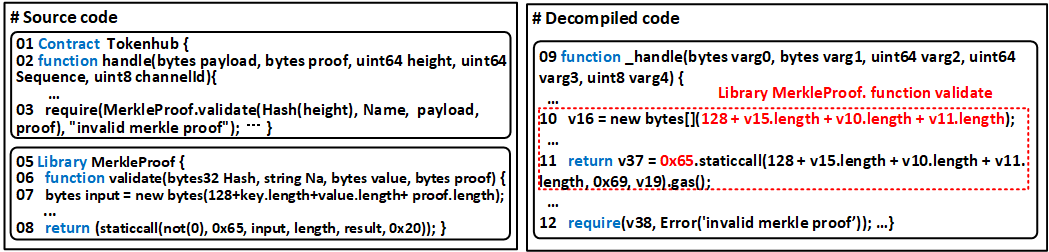}
\caption{An example to show that state-of-the-art decompiler (i.e., Gigahorse~\cite{grech2019Gigahorse}) misses the critical inherited methods.} 
\label{fig:inheritance}

\end{figure*}

\para{C1: Recovering the inherited methods for smart contract}
As mentioned earlier, inheritance is one of the main mechanisms that smart contracts utilize subcontracts, 
making the identification of inherited methods critical for \bugname{} detection.
However, while the inheritance between the main contract and base subcontracts are explicitly stated in the source code level, they are missed at the bytecode level in smart contracts. Specifically, when contract $A$ inherits from the other base subcontracts $B_1, B_2,...,B_n$, all base subcontracts are compiled as code blocks without retaining class information and the virtual function table, for embedding within the contract $A$. 
Figure~\ref{fig:inheritance} presents an example that SOTA decompiler (i.e., Gigahorse~\cite{grech2019Gigahorse}) misses the critical inherited methods.
While the inherited method \textit{validate} is explicitly declared in the source code (line 06-08), it is missing in the bytecode and uncovered by decompiler. By checking the decompiled code carefully, we found that method \textit{validate} is embedded into method \textit{\_handle}. 
Obviously,  Gigahorse~\cite{grech2019Gigahorse} struggles with such embedding, results in failing to recover the inherited method structure. 
While prior research~\cite{liao2025augmenting} optimizes method boundaries at the decompiled code level, it is actually limited by the decompilation errors that are introduced by the decompiler into the decompiled code, results in false positives and false negatives.
In addition, in traditional language such as C++ and Java, prior notable efforts~\cite{wu2023libscan, qiu2021libcapsule} rely on class information and virtual function tables to reconstruct inheritance structures, 
so they are infeasible for recovering the inherited methods for smart contract.


To overcome this challenge, \system{} utilizes transfer learning to effectively recover the inherited methods from the bytecode of smart contract, by learning the pattern for the border (i.e., start point and end point) of inherited methods from training corpus (See Section~\ref{sec: RecoveryforInheritance}). Here, \system{} utilizes the key insight that the context for the border of inherited functions is highly similar between different smart contracts and has generic patterns, regardless of the level of their representations (i.e., source code or bytecode). We take Figure~\ref{fig:inheritance} again for illustration, a method generally declares local variables near the start point (line 10) and utilizes the keyword \textit{return} near the endpoint (line 11).
To this end, \system{} trains a transfer learning model, which effectively learns the generic patterns for the border of inherited methods according to its context. With the transfer learning model, given the intermediate representation (IR) decompiled from contract bytecode, \system{} can effectively recover the inherited functions for smart contracts.


Further, \system{} captures features throughout the methods and establishes an accurate method-level similarity comparison for identifying the usages of subcontract in the given contract.
Previous studies~\cite{zhang2019libid, wu2023libscan} focus on traditional languages such as Java and C++, extract the limited features (e.g., method call, field, control flow) for identifying code reuse, which are not sufficient for \bugname{} identification. 
As countermeasures, our work complements and extends previous research by considering four new types of features, including new mechanism, message call, event, and pre-compiled methods (See Section~\ref{sec: SignatureGeneration}), allowing \system{} to conduct an accurate similarity comparison between methods.
Subsequently, \system{} generates a signature for methods based on the extracted feature. Given a smart contract, \system{} compares the method signatures in pair between the contract and corresponding official subcontract, to identify the usage of subcontract in the given smart contract.




\para{C2: \bugname{} identification} 
The last challenge lies in how to accurately identify \bugname{} based on the identified subcontract usages.
First, identifying \bugname{} relies on extracting apriori knowledge of subcontract, such as access control specifications and conflicts of subcontracts. We take the example of Figure~\ref{motivatingexample} as instances for illustration. 
For example, by reviewing official documentation~\cite{Solidity}, the method \textit{\_xp} of Library \textit{SwapUtils} has overlapping functionality with method \textit{\_xp} of Library \textit{MetaSwapUtils}. A smart contract should carefully check the parameters passed by the external call when reusing the Library \textit{Address}.
Second, identifying \bugname{} relies on fine-grained control flow and data flow analysis. Turning to the example shown in Figure~\ref{motivatingexample}. 
As discussed earlier, detecting SMV for instance of Figure~\ref{motivatingexample} (a) depends on identifying that the functionality-overlap methods affect the same state variable \textit{balance}. And detecting SMV for instance of Figure~\ref{motivatingexample} (b) depends on locating the control-flow pattern of omitting security checks.


To address this challenge, 
\system{} derives access control specifications of subcontracts from official platforms (i.e., OpenZeppelin~\cite{OpenZeppelin}, Node Package Manager (NPM~\cite{NPM}) and obtains the subcontract conflicts from both these official sources and public attack reports, for constructing a priori knowledge database. Subsequently, \system{} highlights the priori knowledge (e.g. subcontract official usages and conflicts) as the SMV indicator  to support vulnerability detection. Finally, \system{} utilizes taint analysis to check if an external contract can be invoked and reaches
state variables with the above SRV indicators. If the condition is satisfied, we consider the analyzed contract vulnerable (See Section~\ref{sec: VulnerabilityIdentification}).



\begin{figure*}[t]
\centering
\includegraphics[width=7in]{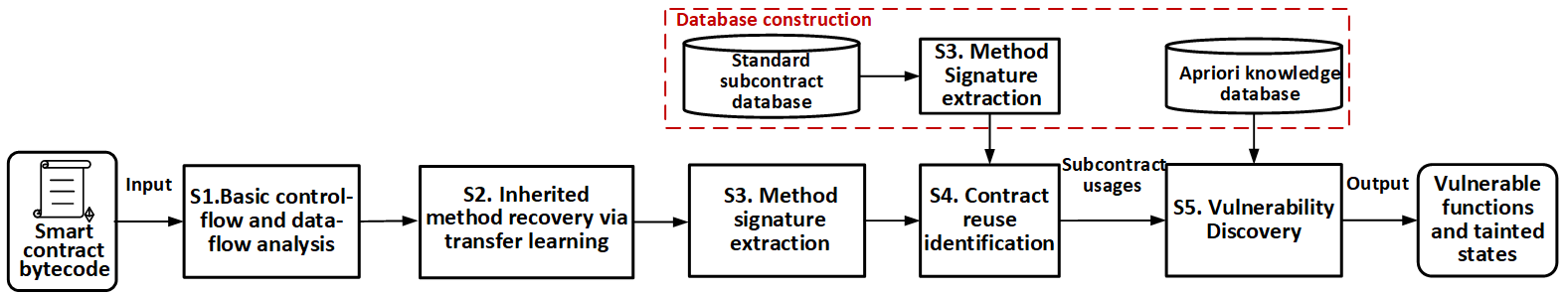}
\caption{The overview of \system{}.} 
\label{fig:overview}

\end{figure*}

\subsection{Workflow of \system{}}
\label{sec: workflow}
\system{} takes the bytecode of the smart contract as input, and ultimately outputs a set of vulnerability traces. Each vulnerability trace is actually a function call-chain that starts from the vulnerable function to tainted state variable(s) affected by external call(s). We present the overview of \system{} in Figure~\ref{fig:overview}. 
As can be seen, the system design of \system{} can be divided into two parts: (1) Database Construction and (2) Workflows for \bugname{} detection.
For database construction, we firstly utilize the largest subcontract dataset  introduced by prior empirical study~\cite{sun2023Demystifying} (i.e., a total of 353 subcontracts), to construct our standard subcontract database which subsequently serves as the basis of similarity comparison.
We then collect official usages for these subcontracts from official
platforms (i.e., OpenZeppelin~\cite{OpenZeppelin}, Node Package Manager (NPM~\cite{NPM}) and obtain the subcontract conflicts from both official platforms and 16 public attack reports, for constructing a priori knowledge database.
In addition, the workflows of \system{} consist of the following steps.


\begin{enumerate}[S1.]
    \item \textbf{Basic control-flow and data-flow analysis.} As a pre-processing step, \system{} utilizes the existing analyzer, SmartDagger~\cite{liao2022SmartDagger}, to generate the control-flow and data-flow graph for the given smart contract.  
    \item \textbf{Inherited methods recovery via transfer learning.} With the control-flow graph, \system{} utilizes the transfer learning model to recover the inherited methods. 
    The output of such a step is the borders of inherited methods, which are critical for \bugname{} detection. 
    \item \textbf{Method Signature extraction.} After recovering the inherited methods, \system{} extracts the features and generates the signatures for methods of the given contracts. In addition, \system{} also performs feature extraction and signature generation for standard subcontracts.
    \item \textbf{Contract reuse identification.} \system{} compares the method signatures in pair between the given smart contract and standard subcontracts. After that, \system{} outputs which version and code snippets of subcontracts are reused in the given smart contract. 
    \item \textbf{Vulnerability discovery.} Based on the identified subcontract usages, \system{} refers to the priori knowledge database for locating the \bugname{} indicators, and subsequently identifies vulnerability traces with taint analysis.
\end{enumerate}

\begin{table}[h]
\footnotesize
\centering
\caption{Notation table.}
\label{tab:notation}
\begin{tabular}{ll}
\toprule
\textbf{Symbol} & \textbf{Meaning} \\
\midrule
$C=\left\{S,E,N \right\}$         & Category labels \\
$P_t$                             & Opcode type similarity score  \\
$F_c$                             & Signature of the contract method $M_c$  \\
$F_s$                             & Signature of the subcontract method $M_s$  \\
$\vec{\mathbf{e_c}}$              & One-hot encoding vector of $F_c$ \\
$\vec{\mathbf{e_s}}$              & One-hot encoding vector of $F_s$ \\
$P_n$                             & Opcode length similarity score  \\

\bottomrule
\end{tabular}
\end{table}

\section{Approach Details}
\label{sec:methodology}






\subsection{Inherited Method Recovery via Transfer Learning} 
\label{sec: RecoveryforInheritance}

After the pre-processing step, \system{} recovers inherited functions for the control-flow graph. Specifically, \system{} regards the recovery of inherited methods as identifying the borders (i.e., start points and end points) of the inherited methods. The key observation here is that, while a smart contract totally omits the information of inherited methods at the bytecode level, but the inherited methods can be recovered according to the context of their borders regardless of the program abstraction level. 

\begin{figure}[t]

\includegraphics[width=3.5in]{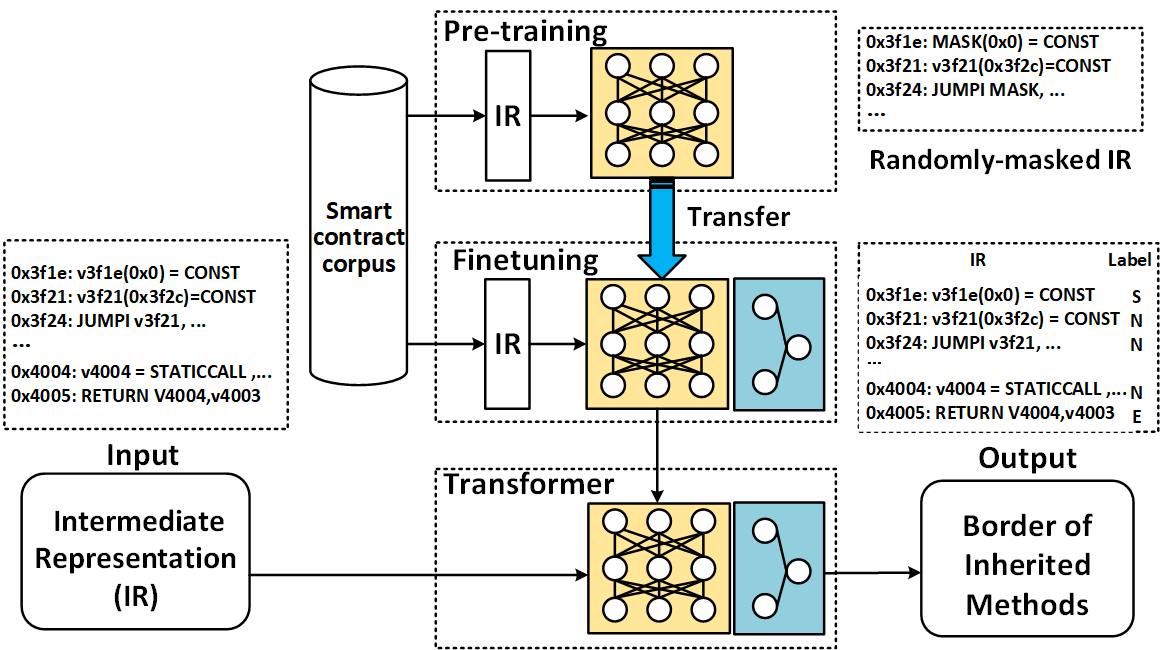}
\caption{Working process of the Transfer Learning model used in \system{}.} 

\label{fig:inheritance_recovery}
\end{figure}

To this end, \system{} utilizes a transfer learning method to recover the inherited functions of smart contracts~\cite{pei2021xda}. Figure~\ref{fig:inheritance_recovery} presents the working process of the transfer learning model. The transfer learning method takes a line of the IR as input, and finally outputs one of the possible category labels $C=\left\{S,E,N \right\}$, where $S$,$E$,$N$ represent start point, end point and neither of them, respectively. The transfer learning method consists of two parts: 1) the transformer model and 2) the training method. The transformer model works as a classifier, dividing each line of the IR into three categories (i.e., start point, end point, and neither of them). In the case of the training method, \system{} teaches the transformer model general dependencies across the IR bytes before teaching it to perform the classification task (identifying the border of inherited functions). In other words, \system{} transfers its learned knowledge of these byte dependencies to the classification task.




\para{Tokenizer design and input representation.} For each input bytecode, Satellite splits it into a sequence consisting of $n$ byte tokens, where each byte has a value ranging from 0x00 to 0xFF. Each byte token is represented as a one-hot encoding vector. In addition to the 256 possible byte values, the input vocabulary includes five reserved tokens including padding $\left \langle PAD \right \rangle$, function start point $\left \langle S \right \rangle$, function end point $\left \langle E \right \rangle$, unknown $\left \langle N \right \rangle$, and mask $\left \langle MASK \right \rangle$. Moreover, we provide no constraints on the size of byte tokens $n$. For instance, the byte token sequence can be either the entire contract bytecode or a subsequence extracted from the contract bytecode. 

\para{Training}
\system{} trains the transformer model via two stages, i.e., pretraining and finetuning, as shown in Figure~\ref{fig:inheritance_recovery}. 

In the first stage, \system{} utilizes mask Language Modeling (mask-LM) to pre-train the transformer model, to teach it the context dependency between IR. 

During the pre-training task, we randomly select 25\% of the bytes within each input token sequence for masking. Among these selected bytes, 50\% are replaced with the special token $\left \langle MASK \right \rangle$, while the remaining 50\% are replaced with random bytes selected from the vocabulary ranging from 0x00 to 0xff. Satellite does not replace all selected bytes with $\left \langle MASK \right \rangle$, because $\left \langle MASK \right \rangle$ token does not exist in subsequent fine-tuning tasks. We aim to prevent the model from learning any spurious semantics from the $\left \langle MASK \right \rangle$ token. Instead, the model should rely on context to predict the masked bytes.

Moreover, we implement a dynamic masking strategy: for the same input token sequence, the set of masked bytes is randomly reselected in different training epochs rather than remaining fixed.


In the second stage, \system{} further performs finetuning on the transformer model, to utilize its comprehension of IR context dependency to address the recovery task of inherited functions. Specifically, \system{} finetunes and updates the parameters of the pretraining transformer model, for minimizing the difference between predicted labels and actual labels, as well as achieve the optimum prediction effect. 

\begin{table*}[b]
\scriptsize
\centering
\caption{Hyper-parameter settings for he transformer model.}
\setlength{\tabcolsep}{1mm}{
\begin{tabular}{l|l|l|l|l}
\hline
\multicolumn{1}{c|}{Type}                                             & \multicolumn{1}{c|}{Size}                                                                                                                  & \multicolumn{1}{c|}{Configuration}                                                                                    & \multicolumn{1}{c|}{Code-Analysis Adaptation}                                                                                                                                            & \multicolumn{1}{c}{Training Protocol}                                                                                                                                                                                                                                                                                                                            \\ \hline
\begin{tabular}[c]{@{}l@{}}RoBERTa \\ transformer\end{tabular} & \begin{tabular}[c]{@{}l@{}}\textbf{Parameters}: 86,160,537; \\ \textbf{Vocabulary size}: 1,000 \\ \textbf{Hidden dimension}: 768; \\ \textbf{FFN dimension}: 3,072\end{tabular} & \begin{tabular}[c]{@{}l@{}}\textbf{Network depth}: 12; \\ \textbf{Attention heads}: 12; \\ \textbf{Activation}: GELU\end{tabular} & \begin{tabular}[c]{@{}l@{}}\textbf{Adaptation layer}: task-specific \\ classification head on  top of \\ RoBERTa\end{tabular} & \begin{tabular}[c]{@{}l@{}}\textbf{Pre-training}: \textbf{Batch size}: 8 seq/GPU; \textbf{Optimizer}: Adam ($\beta_1, \beta_2= 0.9,0.98$); \\ \textbf{Learning rate}: $1\times10^{-4}$ (10,000-step warmup); \textbf{Training steps}: 305,000; \\ \textbf{Fine-tuning}: \textbf{Learning rate}: $1\times10^{-5}$ (500-step warmup); \textbf{Epochs}: 100; \\ \textbf{Batch size}: 8 seq/GPU; \textbf{Optimization objective}: accuracy maximization\end{tabular} \\ \hline
\end{tabular}
}
\label{table.feature}
\end{table*}
\begin{table*}[t]
\footnotesize
\centering
\caption{Identification rule for intra-method opcode feature.}
\begin{tabular}{cc|l|l}
\hline
\multicolumn{2}{c|}{Feature}                                                                             & \multicolumn{1}{c|}{Opcode}                                                                                       & \multicolumn{1}{c}{Symbol} \\ \hline
\multicolumn{1}{c|}{\multirow{2}{*}{Field}} & Read                                                       & MLOAD, SLOAD                                                                                                      & R                          \\ \cline{2-4} 
\multicolumn{1}{c|}{}                       & Write                                                      & MSTORE, SSTORE                                                                                                    & W                          \\ \hline
\multicolumn{1}{c|}{\multirow{2}{*}{Call}}  & \begin{tabular}[c]{@{}c@{}}Method  call\end{tabular}     & \begin{tabular}[c]{@{}l@{}}JUMP, JUMPI, JUMPDEST, CALL, STATICCALL, DELEGATECALL\end{tabular}                  & C0                         \\ \cline{2-4} 
\multicolumn{1}{c|}{}                       & \begin{tabular}[c]{@{}c@{}}Message  call\end{tabular}    & \begin{tabular}[c]{@{}l@{}}CALLER, CALLDATASIZE, CALLDATALOAD, CALLVALUE, CALLDATACOPY, CALLCODE\end{tabular} & C1                         \\ \hline
\multicolumn{2}{c|}{If}                                                                                  & LT, GT, SLT, SGT,EQ, ISZERO                                                                                       & I                          \\ \hline
\multicolumn{2}{c|}{Return}                                                                              & \begin{tabular}[c]{@{}l@{}}RETURNDATASIZE, RETURN RETURNDATACOPY,RETURNPRIVATE\end{tabular}                     & Re                          \\ \hline
\multicolumn{2}{c|}{EVENT}                                                                               & LOG0-LOG4                                                                                                         & E0                         \\ \hline
\multicolumn{1}{c|}{Mechanism}              & \begin{tabular}[c]{@{}c@{}}State-reverting\end{tabular} & REVERT                                                                                                            & M1                         \\ \hline
\multicolumn{1}{c|}{}                       & GAS                                                        & GAS, GASPRICE, GASLIMIT                                                                                           & M2                         \\ \hline
\multicolumn{2}{c|}{Pre-compiled method and contract}                                                               & 0x01-0x09                                                                                                         & P1-P9                      \\ \hline
\end{tabular}
\label{table.intrafeature}
\end{table*}

\begin{figure}[tbp]
\centering
\includegraphics[width=3in]{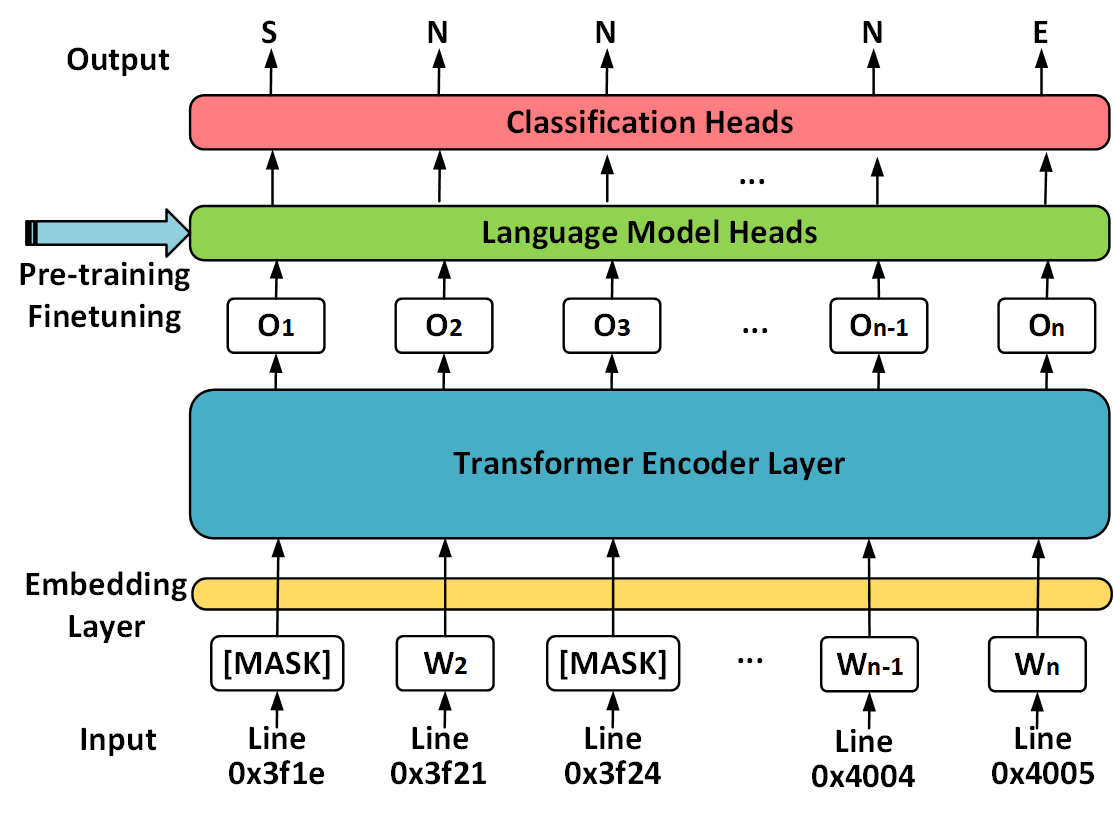}
\caption{Detailed design of the transformer model.} 

\label{fig:transformer}
\vspace{-1.61mm}
\end{figure}

\para{Transformer model}
As shown in Figure~\ref{fig:transformer}. The transformer model is composed of, (1) an embedding layer that converts the input (i.e., a line of IR) into vectors (e.g., $W_2, W_n$) for embedding into the transformer encoder layer; (2) a transformer encoder layer which encodes the vectors and learns the context information for IR, finally outputs the context information (e.g., $O_1, O_2,..., O_n$) into language model heads; (3) language model heads which utilize the learned knowledge from pre-training to process the context information, to facilitate the classification task of classification heads; (4) classification heads which classify each line of IR as one of the possible category labels.


In addition, \system{} performs transfer learning with our constructed corpus. Among the training corpus, each entry of the training data consists of two parts: (1) Smart contract IR and (2) a manually-labeled look-up table that maps each line of IR with one of the possible category labels (i.e., $C=\left\{ S,E,N \right\}$). The details for training corpus construction are further illustrated in Section~\ref{sec: Setup}.

\subsection{Method Signature Extraction} 
\label{sec: SignatureGeneration}

Since contract bytecode does not contain class information, \system{} extracts the features and performs similarity comparison for code reuse identification on the method level. In this step, based on IR that has been recovered inherited methods, \system{} separately extracts the feature for methods of given smart contract and subcontract, as well as generates corresponding signatures. Existing source code-level studies convert
source codes into token streams for signature generation~\cite{chen2021Understanding}, but contract bytecode is without such information. Therefore, \system{} considers extracting features from method opcode (i.e., the critical part of bytecode) for signature generation. Specifically, \system{} considers the following two types of features while generating signatures.


\para{Intra-method opcode feature}
The challenge lies in how to extract as many syntax features and semantic features as possible from the original bytecode. To this end, \system{} utilizes the features shown in Table~\ref{table.intrafeature} to construct the textual representation for intra-method opcode. Specifically, \system{} records the following critical information in order.
\begin{itemize}
     \item \textit{Field}. Such a feature reflects the read and write on variables in the method.  
    \item \textit{Call}. Unlike previous research on JAVA~\cite{zhang2019libid}, \system{} not only considers method call between contracts, but also considers unique message call of smart contracts (i.e., invocation that user calls contract). 
    \item \textit{Syntax}. \system{} mainly considers two types of critical syntax features, (1) conditional branch feature (i.e., \textit{if}) and (2) return statement feature (i.e., \textit{return}).
    \item \textit{Mechanism}. Particularly, \system{} considers two unique mechanisms, i.e., the State-reverting mechanism and Gas mechanism. The former one refers to the mechanism used for error handling and access control, and represents the critical feature of access control in smart contracts. And Gas is the fuel for executing smart contracts and is paid by the user who invokes the contract, so Gas mechanism represents the critical feature of contract execution.
    \item \textit{Event}. The event is used to record the critical log of smart contracts. \system{} considers the event feature, because the event is generally utilized as critical information for interaction between on-chain smart contracts and off-chain front-end applications. For example, the front-end part of DApp monitors the event of the smart contract to observe the critical information (e.g., balance) for users.
    \item \textit{Pre-compiled method}. The virtual machine of the blockchain contains pre-compiled methods that provide frequently used and key functionality such as verification. 
\end{itemize}





\para{Call-chain opcode feature}
To this end, \system{} extracts the call-chain opcode feature to represent the inter-method behavior. Specifically, for a given method, \system{} computes call chains for it, and generates call-chain opcode features by orderly connecting intra-method opcode features of all methods that lie in the call chains.

\begin{figure}[h]
\centering
\includegraphics[width=3.2in]{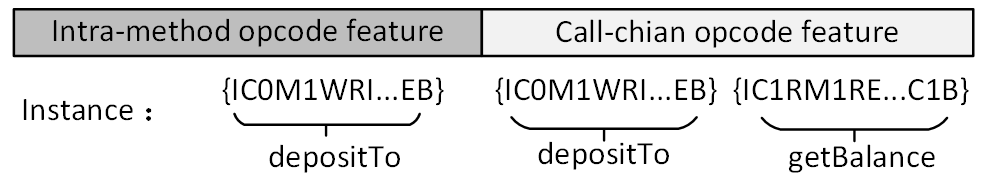}
\vspace{-3mm}
\caption{Format of a method signature with an instance of method \textit{depositTo}.} 
\label{fig:signature}
\end{figure}

We take the method \textit{depositTo} of Figure~\ref{motivatingexample} as an instance to show the signature generation, and the generated signature is shown in Figure~\ref{fig:signature}. As can be seen, the signature is composed of two parts, i.e., intra-method opcode feature and call-chain opcode feature. Specifically, the intra-method opcode feature orderly utilizes symbols to record various features encountered in the program, and the call-chain opcode feature orderly connects intra-method opcode features of methods \textit{depositTo} and \textit{getBalance} according to the call chain (i.e., $depositTo \rightarrow getBalance$).


\subsection{Contract Reuse Identification} 
\label{sec: ReusageIdentification}

For the given contract, \system{} identifies the code reuse by comparing the signature similarity of methods between the contract and subcontract in pair. For example, if \system{} identifies method $M_c$ of contract $C$ is similar to method $M_s$ of subcontract $S$ in terms of signature, \system{} output that contract $C$ reuse method $M_s$ of subcontract $S$. Similar to prior research~\cite{wu2023libscan}, \system{} identifies the similarity of signatures between two methods by comparing the type and number of opcodes within their signatures.

\para{Opcode type similarity}
Firstly, \system{} computes a one-hot encoding vector $\vec{\mathbf{e}}$ for each of the two method signatures. Then, \system{} utilize the cosine similarity to compute the opcode type similarity score $P_t$, because cosine similarity is better suited for assessing the alignment of vectors. 

\begin{small}
\vspace{-0.05in}
\begin{equation}
    P_t= \frac{ \vec{\mathbf{e_c}} \cdot \vec{\mathbf{e_s}} }{\|\mathbf{e_c}\| \cdot \|\mathbf{e_s}\| } 
\end{equation}
\vspace{-0.03in}
\end{small}
where $\vec{\mathbf{e_c}}$ represents the one-hot encoding vector corresponding to the signature of the contract method $M_c$, and $\vec{\mathbf{e_s}}$ represents the one-hot encoding vector corresponding to the signature of subcontract method $M_s$. A high similarity score $P_t$ indicates that two signatures are similar in terms of opcode type. \system{} is configured with a similarity-score threshold $\theta_1$, only if $P_t \ge \theta_1$, \system{} consider two signatures are probably similar in terms of opcode type.


\para{Opcode length similarity}
Further, \system{} computes another similarity score $P_n$ in terms of the number of opcode.
\begin{small}
\vspace{-0.05in}
\begin{equation}
    P_n=\frac{sizeof \left ( F_s \right ) }{sizeof \left ( F_c \right ) } 
\end{equation}
\vspace{-0.03in}
\end{small}
where $F_s$ represents the signature of method $M_s$, and $F_c$ represents that of method $M_c$. $sizeof \left ( \right )$ is the function that gets the size of method signature. Similarly, A high $P_n$ indicates that  two signatures are similar in terms of opcode number. We also configured \system{} with another similarity-score threshold $\theta_2$, only if $P_n \ge \theta_2$, \system{} considers two signatures are probably similar in terms of opcode number.

\vspace{-1mm}
\subsection{Vulnerability Discovery} 
\label{sec: VulnerabilityIdentification}



\para{\bugname{} indicator}
Unlike previous work~\cite{chen2021Understanding} that detects \bugname{}  by conducting the similarity comparison between a given smart contract and known vulnerable subcontract, \system{} takes a set of fine-grained rules for locating vulnerable methods according to the three types of \bugname{}. More specifically,
for each method in smart contracts, \system{} identify it as an \bugname{} indicator (i.e., vulnerable method) only if the identification rules (instance) summarized in Table~\ref{table-indicatorRule} are satisfied.

\begin{table}[b]
    \scriptsize
    \centering
    \renewcommand{\arraystretch}{1.3}
    \caption{Vulnerability indicator rules for SMV detection}

\begin{tabular}{|l|l|}
\hline
Vulnerability subtype              & Vulnerability indicator rule                                                                                                                        \\ \hline
Conflict between $S_{1} \& S_2$ & \begin{tabular}[c]{@{}l@{}}$(isFunctionallySimilar(S_1,S_2))\wedge (\exists v_{s},$ \\ $s.t., Writeon(S_1,v_{s}), Writeon(S_2,v_{s}))$\end{tabular} \\ \hline
Lack of security check on $S$             & $(isLackof(a_{cc}))\wedge (isModified( v_{s}))$                                                                                                     \\ \hline
\end{tabular}    
    \vspace{0.06in}
    
    {$S$  means the subcontract, $v_{s}$  represents state variable, $DB_{vul}$ refers to vulnerability database.}

\label{table-indicatorRule}
\end{table}


While \system{} identifies vulnerable methods well based on the above identification rules, \system{} still needs to satisfy the following condition for locating \bugname{}.

\textbf{Condition-1. Identifying the entry trace for the external call.} On the control flow graph, \system{} searches for the external entry trace for each \bugname{} indicator. Such search is modeled as a reachability analysis process that \system{} starts from the entry points of a given smart contract (i.e., public method) and further checks whether the external call can reach \bugname{} indicators alongside the control-flow paths.

\textbf{Condition-2. Exploring state variables affected by \bugname{}.} After finding the entry trace for the external adversary, \system{} continues to perform taint analysis on the data-flow graph, for finding state variables affected by \bugname{}. Here, identifying affected state variables is to present the undesirable effect of \bugname{} intuitively. Finally, \system{} outputs the vulnerable functions and affected state variables as vulnerability traces which reveal how an external adversary exploits the \bugname{}.




We utilize the motivating example in Figure~\ref{motivatingexample} (b) as an instance to describe the process for vulnerability discovery. Specifically, \system{} identifies method \textit{\_swap} as \bugname{} indicator, as method \textit{\_swap} omits access control on argument \textit{data} (line 03). Then, \system{} screens for condition-2, and locates method \textit{\_swap} as the entry point of execution paths, because method \textit{\_swap} is a public method that can be reached by external call (line 02). Lastly, \system{} locates the state variables affected by \bugname{} by investigating condition-3 (line 09). Therefore, \system{} reports to contract \textit{Swap} with a vulnerable trace (i.e., $ \_swap \rightarrow CallWithValue \rightarrow \left\{ ETH balance \right\} $).



\begin{figure}[t]
\centering
\includegraphics[width=3.5in]{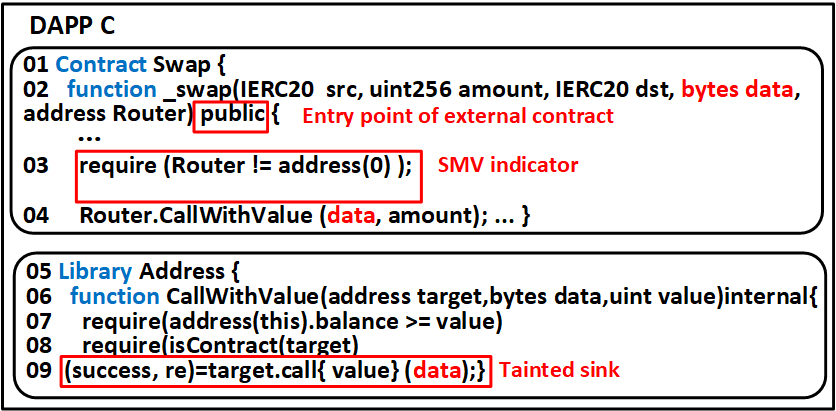}


\caption{The process of \bugname{} detection for example in Figure~\ref{motivatingexample} (b).}
\label{detection}
\end{figure}

\section{Evaluation}
\label{sec:eval}



\subsection{Implementation and Evaluation Setup}
\label{sec: Setup}

We implement \system{} under the environment of Python 3.8.10, which has the tool dependency on SmartDagger~\cite{liao2022SmartDagger}. All the evaluation experiments are conducted for \system{} on a Ubuntu 20.04 server with an Intel i9- 10980XE CPU (3.0GHz), an RTX3090 GPU and 250GB RAM. 

\para{Dataset and ground-truth}
We utilize the following two datasets for evaluation experiments:

\textit{Manually-labeled real-attack \bugname{} dataset $D_m$.} We establish the ground truth for this dataset to evaluate the effectiveness of \system{} and the impact of its each component.
To construct this dataset, we collect vulnerable decentralized applications (DApps) based on 16 real-world \bugname{} attacks. However, only 15 vulnerable DApps, encompassing a total of 981 Solidity documents, were available for analysis. 
Further, by meticulously reviewing these \bugname{} attacks, we manually label the \bugname{}s within the corresponding smart contracts. Specifically, each \bugname{} is composed of two parts, i.e., (1) a vulnerable smart contract and (2) vulnerable traces that lie in the smart contract and trigger the vulnerability. 
Through this labeling process, we annotate a total of 58 \bugname{} traces within the 15 vulnerable DApps. 
To avoid the basis, our four domain experts (i.e., researchers with
three years of experience in smart contract) conduct the manual audit and annotation process using a majority voting approach.  
To the best of our knowledge, $D_m$ is the most comprehensive \bugname attack dataset derived from public sources.


\textit{Library-misuse dataset $D_b$.} 
To enhance the diversity of SMV patterns in the evaluation of Satellite, we further try our best to collect additional SMV patterns from SOTA studies and open-source platforms (e.g., GitHub). We found that the research relevant to Satellite's scope is a prior empirical study~\cite{huang2021Hunting}, which constructs a library-misuse dataset based on 500 audit reports collected from the official websites of five audit companies. From this dataset, we extract 56 contracts containing SMVs, among which there are 4 variable-conflict SMVs and 52  access-control-incompleteness SMVs. To the best of our knowledge, these represents the most comprehensive publicly available SMV patterns from academic and open-source platforms.

\begin{table}[h]
\footnotesize
\centering
\caption{Distribution for the Datasets ($D_{m}$ and $D_{b}$).}
\begin{tabular}{l|cc|cc}
\hline
Dataset & \multicolumn{2}{c|}{Lack of security check} & \multicolumn{2}{c}{Variable conflict} \\ \hline
        & \# Instance            & Rate               & \# Instances         & Rate           \\ \hline
Dm      & 27                     & 23.68\%            & 31                   & 27.19\%        \\
Db      & 5                      & 4.39\%             & 51                   & 44.74\%        \\ \hline
Total   & 32                     & 28.07\%            & 82                   & 71.93\%        \\ \hline
\end{tabular}

\label{table.Distribution}
\end{table}

\textit{Large-scale contract dataset $D_l$.}
To evaluate the effectiveness of \system{} in the wild, 
we utilize the largest available open-source contract dataset~\cite{xblockcontract}, comprising a total of 963,151 smart contracts as of April 22, 2024, referred to as our large-scale contract dataset $D_l$.
As a typical pre-processing process, 
 we perform the deduplication on $D_l$, resulting in 400,445 unique smart contracts. From this deduplicated set, we randomly selected 10\% of these contracts (i.e., 40,044 contracts) for evaluation. Among these contracts, 75\% of them (i.e., 30,033 contracts) were used to construct the training corpus for transfer learning, while the remaining 25\%  (i.e., 10,011 contracts) were utilized to evaluate \system{}'s ability to identify \bugname{}s in the real world.


\para{Evaluation metrics}
In our evaluation, we propose four research questions (RQs) as follow.

\begin{enumerate} [RQ1.]
    \item How effective is \system{} in terms of detecting \bugname{}?
    \item How effective are individual components of \system{} in terms of helping \system{} improve
the precision and recall?
    \item How does \system{} perform compared to other state-of the-art mechanisms in terms of detecting \bugname{}?
    
    \item How do features and thresholds affect the overall performance of \system{}?
    \item What is time complexity of \system{}?
    
    \item Can \system{} detect new \bugname{} from real-world smart contracts?
\end{enumerate}





\begin{table}[h]
\footnotesize
\centering
\caption{Overall effectiveness for \system{} on the Dataset ($D_{m}$), which indicates Satellite’s good
performance in identifying SMVs within smart contract.}

\begin{tabular}{cl|ccc|ccc}
\hline
\multicolumn{2}{c|}{Metrics }                                                    & \multicolumn{3}{c|}{Precision}                                      & \multicolumn{3}{c}{Recall}                                         \\ \hline
\multicolumn{2}{c|}{}                                                                                               & TP                   & FP                   & Rate                  & TP                   & FN                   & Rate                 \\ \hline
\multicolumn{1}{c|}{\multirow{2}{*}{Variable conflict}}                                                        & $D_m$ & 28                   & 6                    & 82.35\%               & 28                   & 3                    & 90.32                \\
\multicolumn{1}{c|}{}                                                                                          & $D_b$ & \multicolumn{1}{l}{4} & \multicolumn{1}{l}{0} & \multicolumn{1}{l|}{100.00\%} & \multicolumn{1}{l}{4} & \multicolumn{1}{l}{1} & \multicolumn{1}{l}{80.00\%} \\ \hline
\multicolumn{1}{c|}{\multirow{2}{*}{\begin{tabular}[c]{@{}c@{}}Lack of security \\check \end{tabular}}} & $D_m$ & 24                   & 4                    & 85.71\%               & 24                   & 3                    & 88.89\%              \\
\multicolumn{1}{c|}{}                                                                                          & $D_b$ & \multicolumn{1}{l}{49} & \multicolumn{1}{l}{9} & \multicolumn{1}{l|}{84.48\%} & \multicolumn{1}{l}{49} & \multicolumn{1}{l}{2} & \multicolumn{1}{l}{96.08\%} \\ \hline
\multicolumn{2}{c|}{Total}                                                                                          & 105                   & 19                   & 84.68\%               & 105                   & 9                    & 92.11\%              \\ \hline
\end{tabular}
\label{table.overall}
\end{table}
\begin{table*}[b]
\footnotesize
\centering
\caption{Comparison results for evaluating the effectiveness of inherited method recovery as well as the effectiveness of method signature extraction, which show that these components  significantly enhances the precision and recall
of Satellite. }
\begin{tabular}{c|ccc|ccc}
\hline
Approach                                                                                       & \multicolumn{3}{c|}{Precision}                                                 & \multicolumn{3}{c}{Recall}                                                    \\ \hline
                                                                                               & TP                     & FP                     & Rate                         & TP                     & FN                     & Rate                        \\ \hline
\begin{tabular}[c]{@{}l@{}}\system{} w\textbackslash{}o inherited  method recovery\end{tabular} & \multicolumn{1}{l}{13} & \multicolumn{1}{l}{9} & \multicolumn{1}{l|}{59.10\%} & \multicolumn{1}{l}{13} & \multicolumn{1}{l}{45} & \multicolumn{1}{l}{22.41\%} \\
\begin{tabular}[c]{@{}l@{}}\system{} w\textbackslash{}o contract-specific feature extraction \end{tabular} & \multicolumn{1}{l}{40} & \multicolumn{1}{l}{8} & \multicolumn{1}{l|}{83.33\%} & \multicolumn{1}{l}{40} & \multicolumn{1}{l}{18} & \multicolumn{1}{l}{68.97\%} \\
\hline
\system{}                                                                                         & 52                     & 10                      & 83.87\%                      & 52                     & 6                     & 89.66\%                     \\ \hline
\end{tabular}
\label{table.inheritance}
\end{table*}

\subsection{Effectiveness of \system{}}
\label{sec: OverallEffectiveness}

To answer RQ1,  
we run \system{} to analyze the smart contracts within the manually-labeled attack dataset $D_m$ and library-misuse dataset $D_b$, and further evaluate its effectiveness through metrics of recall and precision.
Specifically, the precision and recall are determined by comparing the results reported by Satellite with the groundtruth of $D_m$ and $D_b$ (i.e., 114 \bugname{}s). Note that no comparative experiments were conducted for \system{}, as \system{} is the first of its kind for detecting \bugname{}s to the best of our knowledge.

Table~\ref{table.overall} presents the evaluation results in terms of precision and recall for \system{}. As can be seen, \system{} has a good performance in terms of different attacks that exploit \bugname{}s. Overall, \system{} achieves a total precision of 84.68\% and a total recall of 92.11\%, which indicates \system{}'s good effectiveness in identifying SMVs within smart contracts. 

\para{False positives and false negatives} To reveal the causes of false positives and false negatives, we conduct a manual review of all erroneous alarms generated by \system{}. 
According to the manual investigation results, most of the 19 false positives stem from limitations of the access control specifications derived from official subcontract platforms.
For instance, \system{} occasionally fails to identify a tiny number of valid access controls due to the limitation of access control specifications, thereby resulting in false positives related to the lack of security check.
For the 9 false negatives, most of them are related to a small number of subcontractors that depend on third-party non-contract packages outside the blockchain.  
Static analysis methods like ours,
are inherently incapable of addressing these issues because these external packages do not appear in the bytecode of smart contracts, which hinders the vulnerability detection and analysis.




\subsection{Ablation Study}
\label{sec: RecoveryImpact}

\para{Impact of Inherited Method Recovery}
To answer RQ2, we first evaluate the impact of the single component of \system{}, i.e., inherited method recovery. As outlined in Section~\ref{sec: RecoveryforInheritance}, the recovery for inherited methods is an important advantage for \system{}, which helps to ensure both soundness and completeness for \bugname{} detection (i.e., eliminating FNs and FPs). For instance, with the recovery for inherited methods, \system{} can identify more reused methods and conduct more accurate similarity comparisons, so as to help to improve the recall and precision of \bugname{} detection. Therefore, the impact of such component in \system{} is reflected in the recall and precision rate. To this end, we conduct the comparison experiment between two approaches, i.e., (1) \system{} and (2) \system{} without inherited method recovery. Specifically, we establish two approaches and run them over the $D_m$ to evaluate their precision and recall.

Table~\ref{table.inheritance} shows the precision and recall of these two approaches. Due to uninstalling the inherited method recovery, the precision of \system{} without inherited method recovery drops to 59.10\%, and its recall drops more rapidly with a recall of 22.41\%. 
These results affirm that inherited method recovery significantly enhances the precision and recall of \system{}.
In addition, we conduct a manual investigation of the false positives and negatives generated by the version of Satellite uninstalling inherited method recovery. 
The investigation results indicate that 43 of 58 false alarms can be resolved with the implementation of our proposed inherited method recovery mechanism. We take the example in Figure~\ref{fig:inheritance} again as an instance for illustration, which has been discussed in Section~\ref{sec:overview} earlier. 
The method \textit{validate} of Library \textit{MekleProof} contains a logic vulnerability of lack of security check.  The intricate details of this vulnerability can be found in prior attack reports~\cite{Tokenhub}. 
Furthermore, contract \textit{Tokenhub} invoke method \textit{validate}, which is consequently exploited the \bugname{} by the adversary. \system{} without inherited method recovery produces false negatives, because it fails to detect the method \textit{validate} which is embedded within the method \textit{handle} at the bytecode level. 
In contrast, \system{} successfully identifies the inherited method \textit{validate} via trandfer learning, thereby mitigating these false negatives.


\para{Impact of Method Signature Extraction}
As discussed in Section~\ref{sec: SignatureGeneration}, another advantage of \system{} is its ability to accurately extract method features that are uniquely pertinent to smart contracts, which notably enhances both precision and recall.
For example, by considering the contract-specific features (e.g., Gas mechanism), \system{} is able to more precisely capture the behavior of methods and perform thorough and accurate similarity comparisons. Consequently, the effectiveness of this component is reflected in the improved precision and recall rates. 

To this end, we conduct a comparative experiment between two configurations: (1) \system{} with full feature extraction, including contract-specific features, and (2) \system{} is configured to extract only traditional features, such as control flow and method calls, excluding contract-specific elements. Both versions are run on the dataset $D_m$ to evaluate and compare their precision and recall outcomes.
Table~\ref{table.inheritance} shows the precision and recall metrics for the two approaches. Due to ignoring the contract-specific features, the precision of \system{} without the extraction of contract-specific features falls to 83.33\%, and its recall decreases more significantly to 65.52\%. These results underscore the contribution of method signature extraction in enhancing both the precision and recall of \system{}. 
Additionally, we manually investigate every false positive and false negative reported by the \system{} version that is configured to extract only traditional features. 
The investigation results reveal that 15 out of 30 false alarms could be effectively addressed by our proposed method signature extraction technique





\begin{table}[h]
\footnotesize
\caption{Summary of SOTA clone detection approaches for solidity smart contracts.}
\centering
\begin{tabular}{l|ccc}
\hline
 Tool Name        & Open-source              & Relevance          & Code-level \\ \hline
 SmartEmbed~\cite{gao2021Checking}       & \ding{51}                & {\color{black!40}\CIRCLE}          & Source code      \\
 EClone~\cite{liu2019enabling}           & \ding{51}                & {\color{black!40}\CIRCLE}          & Bytecode      \\
 ZepScope~\cite{liu2024using}         & \ding{51}                & \LEFTcircle          & Source code   \\
 SmartSD~\cite{tian2022ethereum}          & \ding{55}                & {\color{black!40}\CIRCLE}          & Bytecode     \\
 VOLCANO~\cite{samreen2022volcano}          & \ding{55}                & {\color{black!40}\CIRCLE}          & Source code     \\ 
 NiCad~\cite{khan2022code}            & \ding{51}                & \Circle          & Source code     \\ 
 \hline
 Satellite        & \ding{51}                & \CIRCLE            & Bytecode     \\ 
 \hline

\end{tabular}

 {\Circle: Lack of support for Vulnerability detection; {\color{black!40}\CIRCLE}: Clone analysis for other vulnerabilities; \LEFTcircle: Partly support for SMV detection;  \CIRCLE: Support for SMV detection;  }
\label{table.studyLLMs}
\end{table}

\begin{table*}[b]
\footnotesize
\centering
\caption{Comparison results for evaluating the impact of contract-specific and additional features, which show that the contract-specific features enhance both precision and recall, whereas the additional features can lead to performance decline.}

\begin{tabular}{l|c|ccccc|ccc}
\hline
Metric    & \multicolumn{1}{l|}{}                                  & \multicolumn{5}{c|}{Contract-specific features}                                                                                                                                                                                                                                                          & \multicolumn{3}{c}{Additional features}                                                                                                                                                         \\ \hline
          & \begin{tabular}[c]{@{}c@{}}Baseline\\ (A)\end{tabular} & \begin{tabular}[c]{@{}c@{}}(A) + MC\\ (B)\end{tabular} & \begin{tabular}[c]{@{}c@{}}(B) + Event\\ (C)\end{tabular} & \begin{tabular}[c]{@{}c@{}}(C) + SR\\ (D)\end{tabular} & \begin{tabular}[c]{@{}c@{}}(D) + PF/C\\ (E)\end{tabular} & \begin{tabular}[c]{@{}c@{}}(E) + Gas\\ (Satellite)\end{tabular} & \begin{tabular}[c]{@{}c@{}}Satellite +\\ Memory\end{tabular} & \begin{tabular}[c]{@{}c@{}}Satellite +\\ Operator\end{tabular} & \begin{tabular}[c]{@{}c@{}}Satellite +\\ Operation\end{tabular} \\ \hline
TP        & 40                                                     & 43                                                     & 43                                                        & 45                                                     & 50                                                       & 52                                                              & 50                                                           & 52                                                             & 49                                                              \\
FP        & 8                                                      & 8                                                      & 10                                                        & 10                                                     & 10                                                       & 10                                                              & 42                                                           & 31                                                             & 38                                                              \\
TN        & 18                                                     & 15                                                     & 15                                                        & 13                                                     & 8                                                        & 6                                                               & 8                                                            & 6                                                              & 9                                                               \\
Precision & 83.33\%                                                & 84.31\%                                                & 81.13\%                                                   & 81.81\%                                                & 83.33\%                                                  & 83.87\%                                                         & 57.61\%                                                      & 62.65\%                                                        & 56.32\%                                                         \\
Recall    & 68.97\%                                                & 74.14\%                                                & 74.14\%                                                   & 77.59\%                                                & 86.20\%                                                  & 89.66\%                                                         & 86.21\%                                                      & 89.66\%                                                        & 84.48\%                                                         \\ \hline
\end{tabular}

\label{table.feature}
\end{table*}

\subsection{Comparison to prior work}
\label{sec: TaintImpact}

As discussed in Section~\ref{sec: VulnerabilityIdentification}, another key advantage of Satellite lies in its indicator analysis, which is instrumental in enhancing the recall of \bugname{} identification.  The effectiveness of vulnerability indicator analysis is particularly reflected in the improved recall rate.

To evaluate the impact of this advantage, we conduct a comparative analysis of \system{}'s recall performance against the SOTA tools. 
Table~\ref{table.studyLLMs} presents the summary of SOTA clone detection approaches for solidity smart contracts~\cite{sun2023Demystifying, liu2024using}. We perform baseline selection according to two criteria, including relevance and public availability. Specifically, we exclude the tools that lack the support for open source or vulnerability detection,
and finally select SmartEmbed~\cite{gao2021Checking}, ZepScope~\cite{liu2024using}, and EClone~\cite{liu2019enabling}. 
Moreover, our selected tools are at both the source-code and bytecode levels (column 3 of Table~\ref{table.studyLLMs}), which facilitates comprehensive and effective  comparison experiments.

Further, since SmartEmbed, ZepScope, and EClone do not support the identification of vulnerable traces (i.e., execution paths) in smart contracts, we specifically compare \system{} with these tools in terms of the number of vulnerable DApps they identify. 
To this end, we ran all the tools over the manually-labeled attack dataset $D_m$ to evaluate their recall rates.


\begin{table}[h]
\footnotesize
\caption{Comparison results between \system{}, EClone, ZepScope and SmartEmbed, which indicate that the indicator analysis component helps Satellite outperform SOTA studies.}
\centering
\begin{tabular}{c|c|c|c}
\hline
Approach   & TP                 & FN                & Recall  \\ \hline
           & \# DAPP (\#Traces) & \# DAPP (\#Traces) &         \\ \hline
SmartEmbed~\cite{gao2021Checking} & 4 (0)               & 11 (0)             & 26.67\% \\

EClone~\cite{liu2019enabling} & 0 (0)               & 15 (0)             & 0.00\% \\

ZepScope~\cite{liu2024using} & 3 (0)               & 12 (0)             & 20.00\% \\

\hline
\system{}     & 13 (52)             & 2 (6)             & 86.67\% \\ \hline
\end{tabular}

\label{table.indicator}
\vspace{-3mm}
\end{table}

As demonstrated in Table~\ref{table.indicator}, \system{} significantly outperforms SmartEmbed, ZepScope and EClone in terms of recall, achieving an impressive rate of 86.67\% and identifying a total of 52 \bugname{} traces. In contrast, the recalls for SmartEmbed, ZepScope and EClone stand at only 26.67\%, 0.00\% and 20.00\%, respectively. These results obviously indicate that the indicator analysis employed by \system{} effectively enhances its recall.
To understand the reasons behind \system{}’s superior performance, we manually inspect all 38 false negatives reported by SmartEmbed, ZepScope and EClone. 
Our analysis reveals that most of these 38 false negatives could be addressed through the vulnerability indicator analysis developed for \system{}.
Again, we take Fig.~\ref{motivatingexample} to show how \system{} avoids
the false negative missed by SmartEmbed and EClone.
In this case, SmartEmbed and EClone are unable to identify SMV, because each individual component of this application (i.e., contract \textit{swap}, Library \textit{SwapUtils} and Library \textit{MetaSwapUtils}) comprises no explicit vulnerabilities that can be detected through code clone analysis. However, \bugname{} is actually caused by a variable conflict arising from the interaction between these components.
By applying the vulnerability indicator rules outlined in Table~\ref{table-indicatorRule}, \system{} is able to identify this interaction and accurately report the \bugname{} traces, thereby eliminating the false negatives. This evidence further substantiates the effectiveness of Satellite’s indicator analysis in improving the \bugname{} detection in smart contracts.




\subsection{Sensitivity Analysis}
\label{sec: sensitivity}






\para{Feature sensitivity analysis}
We evaluate the impact of each contract-specific feature on the effectiveness of Satellite, and explore whether additional features further enhance its performance. Note that we did not evaluate the effectiveness of the non-contract-specific features(rows 2–6 of Table~\ref{table.intrafeature}), because their effectiveness has already been extensively analyzed in prior studies~\cite{sun2023Demystifying,wu2023libscan,qiu2021libcapsule}.

To this end, we consider five contract-specific features including \textit{Message calls (MC), \textit{Events}, \textit{State-reverting (SR)}, \textit{Gas}}, and \textit{Pre-compiled functions/contracts (PF/C)}, and three additional features involving \textit{Operation} (e.g., binary, unary), \textit{Memory} (e.g., array), and \textit{Operator} (e.g., PUSH). Noting that these three additional features represent the most of features that are not captured by Satellite~\cite{huang2021Hunting}.

For the experiments, we first implement a baseline version of Satellite that omits all contract-specific features. We then implement multiple versions of Satellite by incrementally adding the contract-specific features and the additional features. We run all versions of Satellite on the manually-labelled real-attack SMV dataset $D_m$, and further measure their precision and recall.

Table~\ref{table.feature} shows the changes to the precision and recall as we incrementally add each contract-specific feature into Satellite. Each contract-specific feature contributes positively to the better performance of Satellite: its recall increases monotonically. For example, the \textit{Pre-compiled functions/contracts (PF/C)} features achieve the largest enhancement, which enhance the recall by 8.61 \%.
By contrast, when introducing the additional features, Satellite's performance degrades. For instance, when the \textit{Operation}, \textit{Memory} and \textit{Operator} features are introduced individually, Satellite's precision decreases by 27.55\%, 21.22\% and 26.26\%. A manual inspection of all false positives indicates that the additional features weaken Satellite’s ability to handle subcontract variants, thereby reducing their precision.

In summary, the contract-specific features adopted by Satellite enhance both its precision and recall, whereas adding the additional features can lead to its performance decline.

\begin{figure*}[t]
    \centering
    \begin{minipage}[t]{.5\textwidth}
        \centering
        \includegraphics[width=0.85\linewidth]{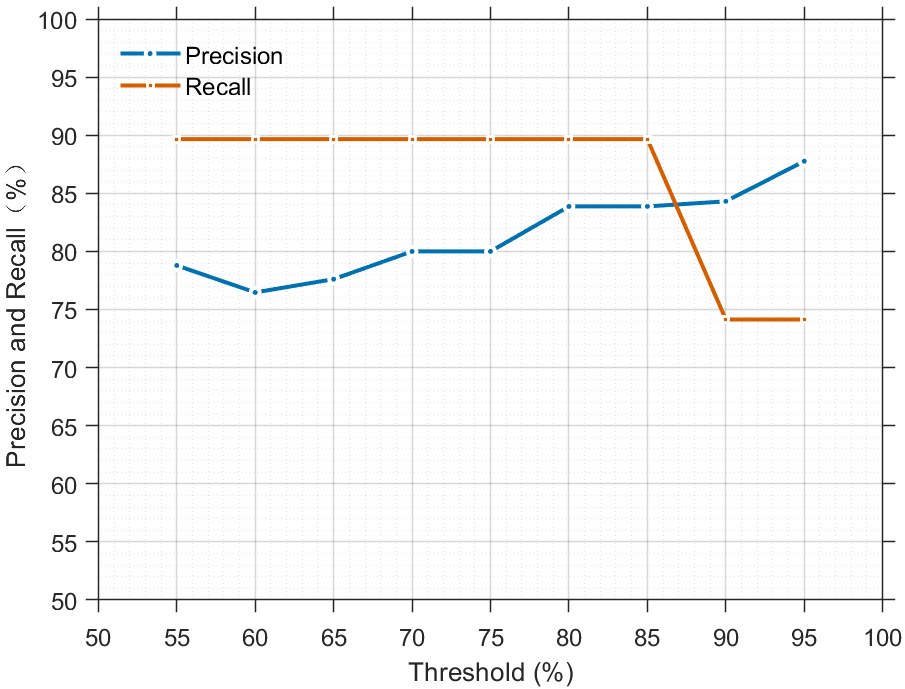}
        \caption{Precision and recall of Satellite under different settings of $\theta_1$.}
        \label{theta1}
    \end{minipage}%
    \begin{minipage}[t]{.5\textwidth}
        \centering
        \includegraphics[width=0.85\linewidth]{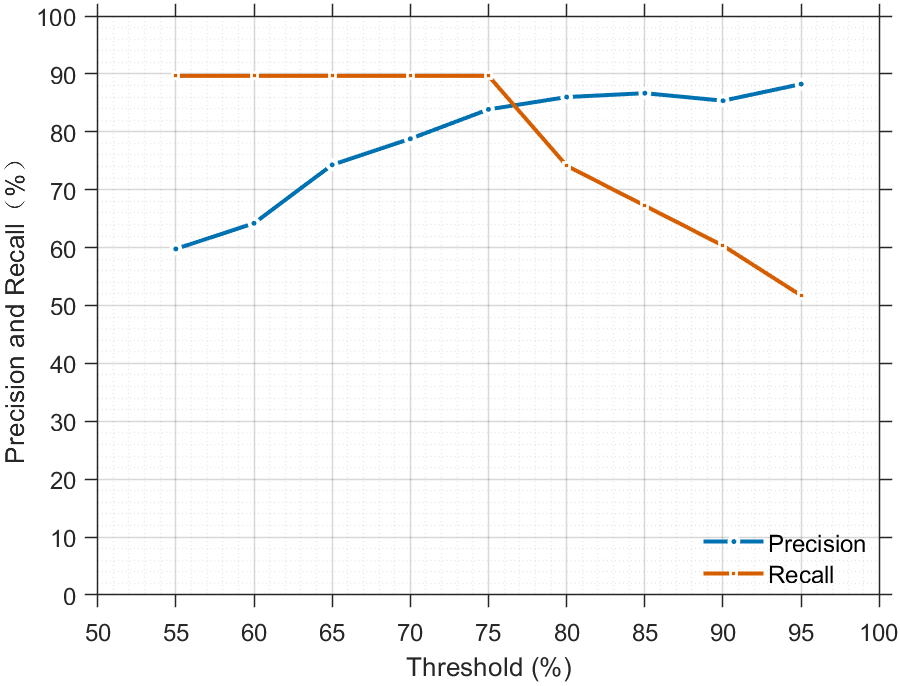}
        \caption{Precision and recall of Satellite under different settings of $\theta_2$.}
        \label{theta2}
    \end{minipage}
\end{figure*}

\para{Threshold sensitivity analysis}
%
%
%
We further evaluat the impact of threshold selection on Satellite’s effectiveness. To this end, we run Satellite with different thresholds for $\theta_1$ and $\theta_2$, and compute the Satellite’s precision and recall. 

Figure~\ref{theta1} presents these two metrics of Satellite under different settings of $\theta_1$. As can be seen, when $\theta_1$ increases, Satellite’s precision increases monotonically, whereas its recall remains stable at first and then decreases gradually. With this observation, we set $\theta_1$ in the range of 80\%–85\% for our evaluation.
Figure~\ref{theta2} shows the precision and recall of Satellite under different settings of $\theta_2$. As can be seen, while $\theta_2$ increases, Satellite’s precision increases monotonically, whereas recall is initially stable and then drops rapidly. Based on this observation, we set $\theta_2$ to 75\% in our evaluation setup.

\subsection{Time complexity}
\label{sec: time}

To address RQ6, we execute Satellite on the large-scale contract dataset $D_l$ and measure its runtime overhead. As shown in Table~\ref{table.efficiency}, the proposed approach can perform SMV detection with an efficient manner. For instance, we compute the average execution times for each individual component of \system{}: 587.51 s for the basic control-flow and data-flow analysis module, 5.48 s for the inherited methods recovery module; 0.10 s for the method signature extraction module, 3.04 s for the contract reuse identification module, and 17.96 s for the vulnerability discovery module.
Noting that the basic control-flow and data-flow analysis module is the most time-consuming component of Satellite, which is caused by the inherent limitations of the third-party analyzer SmartDagger. 
\system{} offers the flexibility to integrate alternative more efficient analyzers  to enhance its efficiency.

In sum, \system{} is proved to be efficient for runtime costs, which indicates its practical value.

\begin{table}[h]
\footnotesize
\caption{The average time for analyzing each contract in the large-scale contract dataset $D_l$.}
\centering
\begin{tabular}{lc}
\hline
Efficiency of \system{} & Avg.time(s)  \\ \hline
Basic control-flow and data-flow analysis     &587.51                        \\ 
Inherited methods recovery     &5.48                       \\ 
Method signature extraction     & 0.10                      \\ 
Contract reuse identification     & 3.04                      \\ 
Vulnerability discovery     & 17.96                      \\ 
\hline
Total     & 614.09                      \\ \hline
\end{tabular}
\label{table.efficiency}
\end{table}

\subsection{Large-scale Analysis for Finding \bugname{}s}
\label{sec: RealWorldPerformance}

To prove that \system{} can detect \bugname{} in the real-world smart contracts, we run \system{} over the random 10,011 smart contracts from the large-scale dataset $D_l$. Our domain experts manually inspect all the results reported by \system{} with a manner of major voting, and finally confirm that \system{} identifies 14 new \bugname{} from 10,011 wild smart contracts. Specifically, \system{} reports 36 warnings (including 30 TPs and 6 FPs
confirmed manually). 16 of 30 \bugname{}s can be detected by SOTA tools (i.e., SmartEmbed~\cite{gao2021Checking}, ZepScope~\cite{liu2024using} and EClone~\cite{liu2019enabling}). Therefore,
\system{} reports 14 (30-16) new \bugname{}. 
Further, by investigating the digital asset affected by these \bugname{}s, we observe that such \bugname{}s affect a total asset of 201,358 USD. Below we
discuss two case studies for illustration.

\para{Case study 1} at \textit{0xeE7477b1C42D173c0791cECF2592F6A8A 4c16b0a.} The function \textit{abort} of
this contract reuses the function \textit{isContract} from the Library \textit{}, which actually contains vulnerabilities. This function \textit{isContract} is intended to implement a Reentrancy guard by restricting external calls from contract accounts. Specifically, its logic assumes that when $extcodesize(addr) > 0$, the address is considered a contract account. In reality, this logic is incorrect. If the address is a contract account within its initialization phase, it also returns 0. Therefore, $extcodesize(addr) > 0$ cannot accurately determine whether an address is a contract account, compromising the contract's reentrancy guard. External attackers can exploit Reentrancy vulnerabilities within the contract to conduct attacks.


\para{Case study 2} at \textit{0x9E2419c8fc0A7C2f2B22CC8dE9AC484aD 00d1F57.} The function \textit{tansfer} of this contract reuse the Library function \textit{Transfer}.
Library function \textit{Transfer}  is an internal method designed for token transfers, which cannot be accessed by external calls and relies on arguments passed by trusted internal calls. Unfortunately, as an external method, method \textit{transfer} overestimates the capacity of Library function \textit{Transfer}, and invokes it directly without implementing a reentrancy guard. This omission in access control introduces an exploitable \bugname{}.  
External attackers can exploit this reentrancy vulnerability to perform malicious actions, potentially compromising the contract.

\para{Responsible disclosure} To avoid undesirable damages (e.g., economic loss) to smart contracts, we anonymize all the results reported by \system{}. In the meantime, we are reaching out the developers of the affected contracts and reporting the identified \bugname{}s. At present, all of the reports are waiting for replies from the authorities such as contract developers or DApp owners.

\section{Discussion and Limitation}
\label{sec: discussion}

\system{} exhibits the following advantages in terms of \bugname{} detection: (1) Evaluation results prove that \system{} is more effective in detecting \bugname{} which are prevalent in the current smart contracts, compared with the SOTA tools. (2) Since \system{} is designed as a bytecode-level analysis framework, such a framework is more practical, especially for contracts that do not disclose their source code to the public. (3) \system{} integrates a set of novel designs, and evaluation results prove that such novel designs help \system{} to reduce false positives and false negatives.
All of the developers, participants, and third-party authorities can utilize \system{} to investigate the security of smart contracts.



\para{Limitation and Future work} A limitation of Satellite is lacking the support for runtime-dynamic SMVs. For example, Satellite suffers performance degradation on a small number of subcontracts that depend on runtime-dynamic non-contract packages outside the blockchain. However, static analysis methods like ours, are inherently incapable of addressing these issues because these external packages do not appear in the bytecode of smart contracts. For future work, we can address this limitation with a dynamic analysis approach that integrates on-chain data (e.g., smart-contract transactions) and off-chain data (e.g., non-contract packages).

Note that Satellite does not support analyzing non-Solidity contracts at present, as its core control-flow analyzer, SmartDagger, is purpose-built for Solidity contracts. Nevertheless, SmartAxe offers the flexibility to integrate alternative analyzers tailored for non-Solidity contracts to extend its capabilities. Crucially, this limitation does not diminish Satellite’s effectiveness for two reasons: (1) The vast majority of smart contracts are written in Solidity, as evidenced by prior research~\cite{liao2023SmartState}. (2) the smart contracts deployed by most DApps exhibit consistent program logic across languages, including both Solidity and non-Solidity. Therefore, the lack of support for non-Solidity contracts does not affect Satellite’s effectiveness in SMV detection.







\para{Threats to validity} 
A construct threat to \system{} is
Satellite’s incomplete capture of the semantics of contract behavior. However, Satellite remains effective in SMV detection for the following reasons: (1) Satellite effectively models the execution, interaction and invocation behaviors via Feature extraction. Satellite’s feature-extraction module incorporates mechanism-related and event-related features (Table~\ref{table.intrafeature}), which effectively capture contract execution behavior (e.g., Gas consumption, state-reverting) as well as front-end/back-end interaction behavior (e.g., events that log the key states of smart contract, which is monitored by DApp front-end). In addition, Satellite considers the call-chain feature which explicitly accounts for contracts’ invocation behavior. (2)
As evidenced by the feature sensitivity experiment (Section V.E), the above-stated behavior features adopted by Satellite effectively enhance both its precision and recall, whereas adding the additional behavior features can lead to its performance decline. In the future, we plan to explore the graph representation to capture more fine-grained semantic dependency within smart contracts to address this limitation.

An internal threat is the manual analysis for the SMV dataset construction, which may introduce subjective bias. To mitigate this threat, we strictly regulate the procedure of manual analysis into following steps. The first step is researcher grouping. We invite six domain researchers with at least two years of experience and divide them into three pairs. Among these, two pairs serve as annotators, while the third pair acts as referee experts who have at least four years of domain research experience.
In the second step, for each pair of annotators, the manual investigation is carried out individually. 
The third step is cross-validation and quality control. For each pair of annotators, they perform the cross-validation and discuss the labels until they obtain a consensus on the classification results. They calculate the Cohen’s Kappa coefficient to measure the consistency of manually-labelled results. If Cohen’s Kappa coefficient falls below the required threshold, we would invite the other pair of referee experts to participate in making the final decision.
In addition, we calculate the average Cohen’s Kappa coefficient to measure the consistency of labeling results, which is 0.758, thus indicating a substantial agreement between each pair of researchers.

An external threat is the lack of support for analyzing non-solidity contracts, as the core control-flow and data-flow analyzer (i.e., SmartDagger) utilized by \system{} is specifically developed for solidity contracts. However, as a generic framework, \system{} provides flexibility, and \system{} can integrate alternative analyzers that are developed for smart contracts written in other languages. Crucially, this limitation does not diminish Satellite’s effectiveness for two reasons: (1) The vast majority of smart contracts are written in Solidity, as evidenced by prior research~\cite{zheng2023dappscan,liao2023SmartState}. (2) the smart contracts deployed by most DApps exhibit consistent program logic across languages, including both Solidity and non-Solidity. Therefore, the lack of support for non-Solidity contracts does not affect Satellite’s effectiveness in SMV detection.



\section{Related Work}
\label{sec:related_work}

\para{Smart Contract Vulnerability Detection}
There has been a number of works aiming at vulnerability detection for smart contracts, which can be categorized as static analysis approaches and dynamic analysis approaches. Oyente~\cite{luu2016Making} is one of the earliest static analysis tools using symbolic execution. And tools such as Slither~\cite{feist2019Slither}, Securify~\cite{tsankov2018Securify},  SmartPulse~\cite{stephens2021SmartPulse}, eTainter~\cite{ghaleb2022ETainterb}, and SmartState~\cite{liao2023SmartState} also leverage static analysis to detect vulnerabilities.  
On the other hand, dynamic analysis tools include ContractFuzzer~\cite{jiang2018ContractFuzzer}, 
Echidna~\cite{grieco2020Echidna},  sFuzz~\cite{nguyen2020SFuzz}, SMARTIAN~\cite{choi2021SMARTIAN}, and ItyFuzz~\cite{shou2023ItyFuzz}. 
However, these tools are not effective in \bugname{} analysis, because they
did not consider the fine-grained feature and context of reuse parts in smart contracts. 

\para{Unexpected Third-Party Code Reuse Detection}
Third-party unexpected code reuse may cause security issues due to the vulnerability propagation~\cite{woo2023V1SCAN} and incorrect usage~\cite{hhe2024code}. 
To mitigate such threats, a range of tools are proposed to discover such vulnerabilities. These tools leverage version information~\cite{zhan2021ATVHUNTER, duan2017Identifying}
or code similarity~\cite{jang2012ReDeBug, xiao2020MVP, kang2022TRACER} to detect potential threats.
While unexpected Code Reuse has become a significant threat, there are a certain number works on this topic for smart contract. Existing works including SmartEmbed~\cite{gao2021Checking}, EClone~\cite{liu2019enabling} VOLCANO~\cite{samreen2022volcano}, NiCad~\cite{khan2022code}  and SmartSD~\cite{tian2022ethereum}, have been dedicated to code clone identification. Other efforts for code clone analysis comprise several empirical studies~\cite{he2020characterizing, pierro2021analysis, khan2022code}. In addition, other notable efforts focus on subcontract misuse issue, including two empirical studies~\cite{sun2023Demystifying, huang2021Hunting} and ZepScope~\cite{liu2024using}.
To the best of our knowledge, \system{} is the first bytecode-level work focusing on detecting subcontract misuse for contracts.     

\section{Conclusion}
\label{sec:conclusion}
This paper proposes a new static analysis framework \system{} for detecting \bugname{} in smart contracts. To evaluate \system{}, we manually annotate a dataset of 58 \bugname{} from real attacks and collect additional 56 SMV patterns from SOTA studies
and open-source platforms (e.g., GitHub). Evaluation results exhibit that \system{} outperforms SOTA tools, with a precision of 84.68\% and recall of 92.11\%. In addition, \system{} effectively detects 14 new/unknown \bugname{}s from 10,011 wild smart contracts, affecting a total amount of digital assets worth 201,358 USD.



\section* {Acknowledgements} 
\label{sec:Acknowledgements}
This research is supported in part by the National Key R\&D Program of China (2023YFB2703600), NSFC/RGC Collaborative Research (62461160332), the National Natural Science Foundation of China (62032025, 62202510, 62572497), Guangdong Zhujiang Talent Program (2023QN10X561) .

\normalem

{
    \bibliographystyle{ieeetr}
    \bibliography{reference}
}




\end{document}